\documentclass[english,russian,12pt,twoside]{article}
\usepackage{amssymb,amsmath}
\usepackage[russian]{babel}
\usepackage[cp866]{inputenc}
\begin{document}

\title{Одна бикватернионная модель
электро-гравимагнитного поля.\\Полевые
аналоги законов Ньютона}
\author{\bf{Алексеева Людмила А.}}
\date{}
\maketitle \centerline{\textit {Институт математики МОН
РК,\,ул.Пушкина, 125,Алма-Ата,050010 Казахстан} }
\centerline{alexeeva@math.kz } \vspace{10mm}

\begin{abstract}
На основе гипотезы об эквивалентности магнитного заряда
гравитационной массе  с использованием инволютивной алгебры бикватернионов
и взаимных кватернионных градиентов разработана математическая модель
электро-грави\-маг\-нит\-ного поля. С введением бикватернионов
напряженности A-поля, заряда-тока, мощности-силы построены уравнения взаимодействия различных
зарядов-токов и на их основе аналоги трех законов Ньютона для
свободных, взаимодействующих полей  и суммарного (единого) поля
взаимодействий.  Получены законы преобразования и сохранения
энергии при взаимодействии полей.
\end{abstract}

Для одной математической модели электро-гравимагнитного
(ЭГМ) поля [1], развивается кватернионный подход
для построения уравнений взаимодействия ЭГМ-полей, порождаемых
различными зарядами и токами. Для  построения модели поля
использовалась гамильтонова форма  симметризованных уравнений
Максвелла [2], которая позволяет легко перейти к
кватернионной записи этих уравнений и законов сохранения [3].
 Комплексификация электромагнитного поля с введением
в уравнения плотности  массы нами названа  \textit{А-полем}.

Здесь, на основе предложенной в [1] гипотезы об эквивалентности
магнитного заряда массе,  разработана
бикватернионная модель
\textit{{\bfА}}-поля. С введением бикватернионов напряженности \textbf{A}-поля,
 заряда-тока, мощ\-нос\-ти-силы полей
 и взаимных комплексных градиентов  построены
 уравнения взаимодействия различных
зарядов-токов и на их основе аналоги трех законов Ньютона для
свободных, взаимодействующих зарядов-токов  и суммарного (единого)
поля взаимодействий.  Получены законы преобразования и сохранения
энергии при их взаимодействии.

Заметим, что бикватернионная запись системы уравнений Максвелла
ранее использовалась рядом авторов для изучения ее решений  и геометрии
 пространства-времени (см., например,[4,5,6]). Здесь для построения
уравнений взаимодействий полей применяем ассоциативную алгебру
бикватернионов с инволюцией и комплексные градиенты, которые несколько
отличаются от ранее
предложенных.\vspace{3mm}

{\bf 1. Гамильтонова форма уравнений Максвелла.} Симметризованные
уравнения Максвелла для A-поля можно записать в виде одного
векторного и одного скалярного уравнения. Здесь запишем
представленные в [2] соотношения в пространстве Минковского ${\bf
R}^{1+3} = \left\{ {(\tau ,x) = (ct,x_1 ,x_2 ,x_3 )} \right\}$:
\begin{equation}\label{(1.1)}
\partial _\tau  A + i\,rotA + J = 0,
\end{equation}
\begin{equation}\label{(1.2)}
\rho  = div\,A =  \rho ^E/\sqrt{\varepsilon}  - i\,
                       \rho ^H /\sqrt{\mu},
\end{equation}
где $A$ - комплексный вектор напряженности:
\begin{equation}\label{(1.3)}
A = A^E  + i\,A^H  = \sqrt \varepsilon \, E + i\,\sqrt \mu \, H,
\end{equation}
$E,H$    -    напряженности   электрического    и гравимагнитного
поля, $\varepsilon,\;\mu $- константы, характеризующие
электрическую и магнитную проводимость и проницаемость среды,  $c
= 1/\sqrt   {\varepsilon  \mu  }  $- скорость  ЭГМ-волн.   $J$   -
ток определяется через электрические и магнитные токи формулой:
\begin{equation}\label{(1.4)}
J = \sqrt {\mu }\, j^E  - i\sqrt \varepsilon \, j^H ,
\end{equation}
а  $\rho  $-  комплексный  заряд А-поля - выражается  через
электрический и магнитный заряды:
\begin{equation}\label{(1.5)}
\rho ^E  = \varepsilon \,div\,E,\quad \rho ^H  =  - \mu \,div\,H.
\end{equation}

\textit{    Замечание.} Если  ввести действительный вектор
скорости $V$ зарядов и считать, что $ j^E  = \rho ^E V,\,\, j^H =
\rho ^H V$, то легко видеть, что
\begin{equation}\label{(1.6)}
J = \rho \bar V,
\end{equation}
где $ \widetilde{V}= V/c$ -- безразмерная скорость.  Однако, представление
(\ref{(1.6)})  не  всегда возможно. Например уравнения допускают
вихревые токи $(div\,J=0)$, для которых $\rho=0$. Если представить
токи в виде суммы потенциальной и вихревой
составляющей:$J=J_p+J_t$, тогда формулы для потенциальной
составляющей электрических и массовых зарядов: $j^E_p  = \rho ^E
V,\,\, j^H_p = \rho ^H V$, -- могут служить для определения их
скорости. Однако эти формулы постулируют параллельность движения
массовых и электрических зарядов и равенство их скоростей, что из
уравнений Максвелла, вообще говоря, не следует. Поэтому формулу
(\ref{(1.6)}) следует воспринимать как эмпирическую для
потенциальных токов ( $rot J =0$).

Энергия  A-поля  $W$ и   вектор   Пойнтинга $P$   при   такой
записи определяются формулами:
$W = 0,5\left( {\varepsilon \left\|
E \right\|^2  + \mu \left\| H \right\|^2 } \right) = 0,5\left\| A
\right\|^2  = 0,5(A,\bar A)$, $P = c^{ - 1} E \times H =
0,5i\,[A,\bar A]$,
 где  $\bar  A = \sqrt \varepsilon  E - i\sqrt
\mu  H$ - комплексно-сопряженное $A$. Здесь всюду $(a,b),[a,b] = a
\times b$ -  скалярное и  векторное произведения $a$ и $b$.

Поскольку квадрат размерности A-поля равен плотности энергии, его
можно назвать \textit{энергетическим}.

В  уравнениях  Максвелла $\rho ^H  = 0$,  т.к.   считается,  что
магнитных  зарядов  в  природе нет.   В  [1]   предложена
гипотеза: \textit{плотность  $\rho  ^H $ эквивалентна плотности
гравитационной  массы.} Если ее принять, то получаем, как покажем далее,
вполне правдоподобные следствия.

Из нее  следует,  что  потенциальная
часть вектора  $H$   описывает гравитационное поле, а вихревая -
магнитное, поэтому  $H$-поле  -  это гравимагнитное   поле.
Следовательно,  $A$-поле   является
\textit{электро-гравимагнитным}.

Будем называть $j^H $ \textit{гравимагнитным }током. При $\rho ^H
= 0$  это  чисто \textit{магнитные} токи, при потенциальном $H$
токи \textit{массовые}. Далее  покажем,  что эта гипотеза имеет
теоретические подтверждения, приводящие к весьма правдоподобным
следствиям.

Приведем здесь также некоторые утверждения для А-поля, доказанные
в [2], которые позволяют обосновать далее кватернионный подход к
построению уравнений A-полей и их взаимодействий.

Т  е  о р е м а  1.1. \textit{При заданных токах и зарядах решение
(\ref{(1.1)}) является решением волнового уравнения}:
$
    \Box \,A  =  (\partial  _\tau  ^2  - \Delta )A = i\,  rot\,J  -
grad\rho  - \partial _\tau  J,    $\\
\textit{и удовлетворяет законам  сохранения заряда и энергии}:
 $
\partial _\tau  \rho  + div\,J = 0$, $\partial _\tau  W + div\,P =  (j^H H - j^E E)/c.  $
\

Из  теоремы 1.1 следуют интегральные соотношения, которые  легко
получить, используя теоремы Остроградского-Гаусса и Стокса.

Пусть  $D^ -$  область в $R^3 $, ограниченная   замкнутой
ляпуновской  поверхностью  $D$  с  единичной  внешней  нормалью
$n(x)$, $S$ - поверхность  с  краем, опирающаяся на контур $l$,
$e_l $- единичный касательный вектор к $l$, $dV(x) = dx_1 dx_2
dx_3 , dD(x),dS(x)$    -    дифференциалы    площади
соответствующих поверхностей, $dl$- дифференциал длины дуги на
$l$.

Т  е  о  р  е  м  а   1.2.  \textit{Решения  уравнений   (\ref{(1.1)})  -
(\ref{(1.2)}) удовлетворяют следующим законам сохранения}:
                                 \[ \begin{array}{*{20}c}
    \int\limits_{D^ -  } {\left\{ {\rho (x,\tau ) - \,\rho (x,0)}
 \right\}} dV(x) +  \int\limits_0^\tau  {d\tau \int\limits_D {(n,J)}
                           \,dS(x)}  = 0, \\
     \int\limits_{D^ -  } {\left( W(x,\tau ) - W(x,0) \right)
 \,dV(x)} + \int\limits_0^\tau  {d\tau \int\limits_D {(n,P)} \,dD(x)}
= \\=\int\limits_0^t {dt\int\limits_{D^ -  } {(j^H H - j^E E)}
\,dV(x)}; \end{array} \]
\[ \begin{array}{*{20}c}
           \int\limits_S {(A(x,\tau ) - A(x,0),n)dS(x)}  +
 i\int\limits_0^\tau  {d\tau \int\limits_l {\left( {A,e_l } \right)}
    dl}  +\int\limits_0^\tau  {d\tau \int\limits_S {\left( {J,n}
                        \right)dS(x)} }  = 0,\end{array} \]
\[ \begin{array}{*{20}c}
 \int\limits_{D^ -  } {\left( {A(x,\tau ) - A(x,0)} \right)\,}
 dV(x) + i\int\limits_0^\tau  {d\tau \int\limits_D {\left[ {n,A}
\right]\,} dD(x)}  + \int\limits_0^\tau  {d\tau \int\limits_{D^
- }{J(x,\tau )} \,dV}  = 0. \end{array} \]

\textit{Ударные волны. }Эти теоремы справедливы также и для
ударных  волн со   скачком   напряженности  поля  на  фронтах,
которые являются обобщенными решениями уравнений Максвелла. Как
показано в  [2],  на фронтах ударных волн выполняются следующие
условия на скачки:
\begin{equation}\label{(1.7)}
\left[ A \right]_{F_t }  + i\,\left[ A \right]_{F_t }  \times m
=0,
\end{equation}
или, для скачков напряженностей:
\begin{equation}\label{(1.8)}
\sqrt{\varepsilon } \left[ E \right]_{F_t }  = \sqrt{\mu }\,
\left[ H \right]_{F_t }  \times m,\quad \sqrt{\mu }\,\left[  H
\right]_{F_t } =-\sqrt{ \varepsilon } \left[ E \right]_{F_t }
\times m.
\end{equation}
Здесь  $m$  волновой  вектор   -  трехмерный  единичный  вектор -
определен  на фронте ударной волны $F_t $,  направлен в сторону
его движения в  $ R^3 $ со скоростью $c$  .

Из  этих соотношений следует, что   $(\left[ E \right]_{F_t  } ,m)
= 0,(\left[ H \right]_{F_t } ,m) = 0$, т.е. ударные волны А-поля
являются \textit{поперечными}.

 Из (\ref{(1.8)})  также следуют  равенства:

\begin{equation}\label{(1.9)}
 \left[ \rho  \right]_{F_\tau  }  = \left( {m,\left[ J
   \right]_{F_\tau  } } \right),\quad
{ [W(x,t)]}_{F_t }  = (m, [P ])_{F_t }.
\end{equation}
Следовательно,   скачок   вектора   Пойнтинга   определяет   как
направление   распространения  ударной   волны,   так   и величину
переносимой ею энергии. Так, если перед фронтом волны $A=0$, то $
m  =  P/\left\|  P  \right\|\,\,(\left\| P \right\| =  \sqrt
{P_1^2  + P_2^2  + P_3^2 } )$, а значения на фронте связаны
соотношениями:
\[ W | _{F_t }  = \| P \|_{F_t},
 \quad \rho  |_{F_t} = (m,J)|_{F_t}.\]

Заметим,  что все соотношения для A-поля (а не для $E$ и  $H$) не
содержат  констант среды,  в частности,  скорость электромагнитных
волн, которая во введенной системе координат безразмерна и равна
1.

В [3] введено новое  представление А-поля, которое кратко опишем в
п.2. \vspace{3mm}

 \textbf{ 2. Бикватернионы А-поля}.    Введем
функциональное пространство   бикватернионов.    Это пространство
комплексных кватернионов:  $\textit{K}({\bf R}^{1+3}) = \{ {\bf  F} =
f(x,\tau ) + F(x,\tau )\} $,  где  $f$ - комплекснозначные функции
$(f = f_1 +  if_2 )$, а $F$ - трехмерная вектор-функция с
комплексными компонентами $(F = F_1  + iF_2 )$,  $f$ и $F$ -
локально интегрируемы  и дифференцируемы на M.   Пространство \textit{К} -
ассоциативная алгебра со сложением вида: ${\bf  F}  +  {\bf  G}  =  (f  +  g)  +
(F  + G)$,    и операцией кватернионного умножения " $ \circ $":
\begin{equation}\label{(2.1)}
{\bf F} \circ {\bf G} = (f + F) \circ (g + G) = (fg - (F,G)) +
                 (fG + gF + [F,G]).
\end{equation}
Здесь $(F,G)=(F_1G_1-F_2G_2)+i(F_1G_2+F_2G_1)$. Квадратные скобки соответствуют обычному векторному умножению.

Бикватернион ${\bf F}^*  = \bar f - \bar F$,  где черта обозначает
соответствующие компонентам комплексно-сопряженные числа,
называется \textit{сопряженным}.   Если   ${\bf   F}^*    =   {\bf
F}$, бикватернион \textit{унитарный}. Таковыми являются бикватернионы вида
$f_1+iF_2$  и только они.

Легко проверяется ассоциативность и свойства
инволютивной алгебры:\\$(\alpha \bf {A})^*=\overline{\alpha} \bf {A}^*,\,
(\bf {A}+\bf {B})^*=\bf {A}^*+\bf {B}^*,\,(\bf{A}\circ\bf {B})^*=\bf {B}^*\circ\bf {A}^*,\,
(\bf {A})^{**}=\bf {A}^*$.

Далее   используем   дифференциальные  операторы   --   взаимные
комплексные градиенты:
\begin{equation}\label{(2.3)}
{\bf D}^ +   = \partial _\tau   + i\nabla ,\quad {\bf D}^ -
                   = \partial _\tau   - i\nabla ,
\end{equation}
где оператор Гамильтона $\nabla  = grad = (\partial _1 ,\partial
_2 ,\partial  _3  )$.  Заметим, что в смысле  выше  данных
определений каждый из них можно назвать \textit{унитарным
оператором}: $({\bf D}^ - )^*  = {\bf D}^ -  ,\;({\bf D}^ +  )^*  =
{\bf D}^ +  $. Их действие на К определено как в алгебре
бикватернионов:
   $${\bf D}^ +  {\bf F} = (\partial _\tau   + i\nabla ) \circ (f +
   F) = (\partial _\tau  f - i\,(\nabla ,F) + \partial _\tau  F +
                      i\nabla f + i[\nabla ,F],$$
   $${\bf D}^ -  {\bf F} = (\partial _\tau   - i\nabla ) \circ (f +
 F) = (\partial _\tau  f + i\,div\,F) + \partial _\tau  F - igrad\,f
                            - i\,rot\,F.$$
Волновой оператор представим в виде: $ \Box= {\bf D}^ -   \circ
{\bf D}^ +   = {\bf D}^ +   \circ {\bf D}^ -  $.

Введем бикватернионы А-поля: потенциал ${\bf \Phi } = i\phi  -
\Psi $, напряженность  $\textbf{A}=0+A$, плотность
энергии-импульса ${\bf \Xi } = W + iP$ , плотность
заряда-тока${\bf \Theta } = i\rho + J$  .

Верны следующие соотношения:
  $${\bf \Xi } = 0,5\,{\bf A}^*  \circ {\bf A} = 0,5\,(\bar A,A) -
                          0,5\,[A,\bar A]$$
      $${\bf D}^ -  {\bf \Phi } = (\partial _\tau   - i\,\nabla
     )(i\phi  - \Psi ) = i(\partial _\tau  \phi  - div\,\Psi ) +
        irot\,\Psi  - \partial _\tau  \Psi  + \,grad\,\phi $$
где ${\bf A}^*  = 0 - \bar A$ - сопряженный кватернион.

Если  потенциал  удовлетворяет лоренцевой калибровке:  $\partial
_\tau  \phi  - div\,\Psi  = 0,$ то
\begin{equation}\label{(2.4)}
{\bf A} = grad\,\phi  -
\partial _\tau  \Psi  + i\,rot\,\Psi,\quad{\bf D}^+  {\bf A} = \, - (i\rho  + J).
\end{equation}
Откуда следует $$\Box{\bf \Phi } =  - {\bf \Theta }\,\,
\Rightarrow \quad \Box\phi  =
                      - \rho ,\,\,\Box\Psi  = J,$$
\begin{equation}\label{(2.5)}
 {\bf D}^ -  {\bf D}^ +  {\bf A} =  - i(\partial _\tau  \rho  +
  div\,J) - \nabla \rho  - \partial _\tau  J + i\,rot\,J = 0 + \Box A
\end{equation}
Из (\ref{(2.5)}), сравнивая кватернионы, следует закон сохранения
заряда и волновое уравнение для А-поля  (см.  теорема 1.1).

Вторая формула (\ref{(2.4)}) -- это симметризованные уравнения
Максвелла, которые фактически определяют заряды-токи как
комплексный градиент А-поля.

Отсюда следует, что просто последовательное (тройное) взятие
комплексных градиентов от потенциала А-поля (с лоренцевой
калибровкой), определяет кватернионы, соответствующие
напряженности поля, зарядам и токам, закону сохранения заряда и
волновому  уравнению для вектора А. Скалярная часть комплексного
градиента кватерниона энергии-импульса А-поля дает закон
сохранения энергии [2].

Итак \emph{заряды и токи - это просто физическое
проявление комплексного градиента напряженности A-поля.
}

Заметим, что подобные закономерности должны наблюдаться для любого
векторного поля с одной скоростью распространения возмущений,
только физическое проявление комплексных градиентов от
кватернионов этого поля будет иметь другую природу. Следует ее
только находить.

 \vspace{3mm}

 \textbf{ 3. Мощность и плотность объемных сил. Третий
закон Ньютона.}    Рассмотрим  два поля ${\bf A}$ и ${\bf A}'$,
${\bf \Theta  }{\bf ,}\,{\bf  \Theta  '}$  -- соответствующие им
(или порождающие их) заряды-токи.  Назовем бикватернион
\begin{equation}\label{(3.1)}
{\bf F} = M - iF =  - \;{\bf \Theta } \circ {\bf A}' =  -
                (i\rho  + J) \circ A' = (A',J) - i\rho A' + [A',J]
\end{equation}
плотностью   \textit{мощности-  силы}, действующей  со  стороны
поля $A'$ на заряды  и  токи  поля $A$ .   Действительно,  с
учетом (\ref{(1.3)}),(\ref{(1.4)}), скалярная часть имеет вид
плотности мощности действующих сил:
\begin{equation}\label{(3.2)}
 M = (A',J) = c^{ - 1} ((E',j^E ) + (H',j^H )) + i((B',j^E ) -
 (D',j^H ))
\end{equation}
Выделяя  действительную  и мнимую части  векторной  составляющей
бикватерниона, получим выражения для плотности объемных сил
$\left( {F = F^H  + i\,F^E } \right)$:

\begin{equation}\label{(3.3)}
Re\, F = \rho ^E E' + \rho ^H H' + j^E
                  \times B' - j^H  \times D' = F^H
\end{equation}
\begin{equation}\label{(3.4)}
Im\, F = \left( {\rho ^E B' - \rho ^H D'} \right) + c^{ - 1}
\left( {E' \times j^E  + H' \times j^H } \right) = F^E
\end{equation}
Здесь  $B  =  \mu H$- аналог вектора магнитной индукции (в
вихревой части совпадает с ним), $ D = \varepsilon E$ -вектор
электрического смещения.

Напряженность  гравитационного  поля  описывается  потенциальной
частью вектора $H$, а роторная часть этого вектора описывает
магнитное поле. Тогда скалярная часть ${\bf \Theta },\,{\bf \Theta
'}$ содержит  плотности  электрического заряда и массы,  а
векторная  - плотности  электрического  тока и тока  массы
(количество  движения массы).

Исходя  из этих предположений, в уравнении (\ref{(3.3)}) стоят
известные массовые  силы,  последовательно: кулоновская  сила,
гравитационная сила,  сила  Лоренца   и  новая $D' \times j^H  $,
которую  назовем \textit{электромассовой}.  В  действительной
части мощности   (\ref{(3.2)})   стоит мощность  кулоновских,
гравитационных  и магнитных  сил.  Мощность электромассовой силы в
действительную часть (\ref{(3.2)})  не  входит,  т.к. она не
работает на перемещениях массы, поскольку перпендикулярна  ее
скорости. Интересно,  что мощность силы Лоренца  в  действительную
часть (\ref{(3.2)}) также не входит, что свидетельствует в пользу
того, что эта сила перпендикулярна скорости массы, хотя
непосредственно из уравнений Максвелла это не следует.

Естественно,  по  аналогии,  предположить, что  уравнения
(\ref{(3.4)}) описывают    силы,   вызывающие   изменение
электрических токов (электрические  силы), а в мнимой части
(\ref{(3.2)}) стоят соответствующие им мощности.

В    силу    третьего   закона   Ньютона   о    действующих    и
противодействующих силах, предположим, что для (\ref{(3.1)}) и
(\ref{(3.2)}) должно выполняться: ${\bf F'} =  - {\bf F}$.
Следовательно, этот закон  для взаимодействующих полей (зарядов и
токов) имеет следующий вид.

\textit{  Закон о действии и противодействии  полей: }
 \[{\bf \Theta } \circ {\bf A'} =  - {\bf \Theta '} \circ {\bf A}.
    \]
Интересно, что в скалярной части он требует равенства плотностей
мощностей соответствующих сил, действующих на заряды и токи
другого поля,  т.е. подобен известному в механике сплошных средств
тождеству взаимности Бетти, которое обычно записывается для работы
сил . \vspace{3mm}

 \textbf{  4. Второй закон Ньютона и уравнения
взаимодействия}.    Поле  зарядов  и   токов меняется под
воздействием  поля  других зарядов  и  токов. Направление наиболее
интенсивного изменения  поля описывает его градиент. По аналогии,
изменение бикватерниона заряда-тока происходит наиболее интенсивно
в "направлении" его комплексного градиента. Естественно
предположить,  что  это  изменение   должно происходить в
"направлении" бикватерниона мощности-силы  со стороны второго поля
на первое.  Поэтому закон изменения заряда-тока поля под действием
другого, подобный второму закону Ньютона, предложим в виде
следующих  уравнений.

\textit{Уравнения взаимодействия зарядов-токов:}
\begin{equation}\label{(4.1)}
\kappa {\bf D}^ -  {\bf \Theta } = {\bf F} \equiv  - {\bf \Theta
 } \circ {\bf A}',\,\,\,\,
\kappa {\bf D}^ -  {\bf \Theta }' =  - {\bf \Theta }' \circ {\bf
A},
\end{equation}
\begin{equation}\label{(4.2)}
 {\bf \Theta } \circ {\bf A}' =  - {\bf \Theta }' \circ {\bf A},
\end{equation}
\begin{equation}\label{(4.3)}
{\bf D}^ +  {\bf A} + {\bf \Theta } = {\bf 0},
                              \,\,\,\,
{\bf D}^ +  {\bf A}' + {\bf \Theta }' = {\bf 0}.
\end{equation}
Здесь   уравнения   (\ref{(4.1)}) соответствуют  второму закону
Ньютона, записанному для зарядов-токов каждого из
взаимодействующих полей, а (\ref{(4.2)})  -  третьему. Вместе с
уравнениями  Максвелла для  этих полей (\ref{(4.3)})  они   дают
замкнутую  систему  нелинейных дифференциальных уравнений  для их
определения.     Введение константы взаимодействия $\kappa $
связано с размерностью.

Раскрывая скалярную и векторную часть (\ref{(4.1)}), запишем

\textit{Уравнения трансформации зарядов-токов А-поля }:
\begin{equation}\label{(4.4)}
 i\,\kappa \left( {\partial _\tau  \rho  + div\,J} \right) = M,
\end{equation}
\begin{equation}\label{(4.5)}
i\,\kappa \left( {\partial _\tau  J - i\,rot\,J + \nabla \rho }
  \right) = F.
\end{equation}
Из  (\ref{(4.4)}),  с  учетом  закона сохранения заряда,  следует,
что при взаимодействии полей $M = 0.$ Отсюда, с учетом
(\ref{(3.2)}) получим
                                 \[
(E',j^E ) + (H',j^H ) = 0,\quad(B',j^E ) - (D',j^H ) = 0
                                 \]
Эти два соотношения можно проверить экспериментально.

Правая часть уравнения (\ref{(4.5)}) соответствует, как мы
обсуждали в п.3, силе.    С учетом (\ref{(3.3)}), (\ref{(3.4)})  и
(\ref{(1.2)}), (\ref{(1.3)}), (\ref{(1.4)}), эти уравнения имеют
следующий вид.

\textit{Аналог второго закона Ньютона для зарядов-токов:}
$$
\kappa  \left(  {\mu  ^{  - 0,5} grad\,\rho  ^H   +  \sqrt
\varepsilon\,\partial _\tau  j^H  + \sqrt \mu  \, rot\,  j^E  }
\right)= $$
\begin{equation}\label{(4.6)}
=\rho ^E E' + \rho ^H H' + j^E  \times B' - j^H \times D',
\end{equation}
 $$
 \kappa  \left(  {\varepsilon ^{ - 0,5}  grad\,\rho  ^E   +  \sqrt
\mu\,\partial _\tau  j^E  - \sqrt \varepsilon  \, rot\,j^H }
\right)=$$
\begin{equation}\label{(4.7)}
= c\left( {\rho ^E B' - \rho ^H D'} \right) + c^{ - 1} \left( {E'
\times j^E  + H' \times j^H } \right).
\end{equation}
Аналогом  количества движения массы здесь в (\ref{(4.6)}) является
$\kappa \sqrt  \varepsilon   j^H  $. Уравнение (\ref{(4.7)})
описывает воздействие внешнего поля на электрические токи, его
аналог автору неизвестен.

Если одно поле намного сильнее второго, например, если   $W'  >
>  W$,  то  можно
изменением  второго поля под воздействием зарядов  и  токов
первого пренебречь.  В этом случае получаем замкнутую систему
уравнений  для определения  движения зарядов и токов первого поля
под  воздействием зарядов и токов второго
\begin{equation}\label{(4.8)}
\kappa {\bf D}^ -  {\bf \Theta } =  - {\bf \Theta } \circ {\bf
A}',
\end{equation}
где   ${\bf  A}'$  известно. Соответствующее им А-поле
определяется уравнениями Максвелла (\ref{(2.4)}) .

\textit{Замечание.} Здесь отметим различие при построении
замкнутой системы уравнений для определения движения зарядов и
токов в сравнении с моделью, предложенной в [1], связанное с тем,
что поле зарядов и токов генерирует силы, воздействующие на
\textit{другие} заряды и токи, а не на самих себя. Поэтому в [1] в
соответствующей системе в качестве напряжений следовало брать
напряжение воздействующего поля. В этом случае правые части
уравнений обеих систем совпадают. Запись левой части системы в [1]
можно рассматривать лишь как первое приближение.

Если   проинтегрировать  соотношение  (\ref{(4.5)})   по
фиксированному объему:
  $$
    \partial  _\tau  \int\limits_{D^ -  } {\kappa \sqrt  \varepsilon\,
j^H  dV(x)  + } \int\limits_{D^ -  } {\kappa \left( {\mu ^{  -
0,5} \,grad\rho  ^H  + \sqrt \mu  \, rot\, j^E } \right)} \,dV(x)
=$$
    $$  =  \int\limits_{D^ -  } {\left( {\rho ^E E' + \rho ^H  H'  +
j^E  \times B' - j^H  \times D'} \right)} \,dV(x),$$
 то получим
формулу изменения количества движения в заданном объеме.

Введем   обозначение  для  массы  и  заряда,  заключенных   в
фиксированном объеме:
                                 \[
m_{D^ -  } (t) = \int\limits_{D^ -  } {\rho ^H (x,t)}
\;dV(x),\quad q_{D^ -  } (t) = \int\limits_{D^ -  } {\rho ^E
(x,t)} dV(x).  \] Используя  теорему Остроградского и полагая, что
подынтегральные функции достаточно гладкие в данной области,
получим:
\[
\kappa  \sqrt  \varepsilon  \;\partial _\tau  \int\limits_{D^  - }
{j^H (x,t)dV(x) }+ \kappa \int\limits_D {\left( n(x)\frac{\rho ^H
}{\sqrt \mu  } + \sqrt \mu \,\left[ n(x),\, j^E (x,t) \right]
\right) \,dS(x)} =
\]
\[
  = q_{D^ -  } \tilde E' + m_{D^ -  } \tilde H' + \int\limits_{D^  -
} {\left( {j^E  \times B' - j^H  \times D'} \right)} \,dV(x).
\]
Здесь в двух первых слагаемых справа стоят значения напряженностей
в некоторой  фиксированной точке данной области (по  теореме
Лагранжа для интеграла от непрерывной функции). Если  $\rho ^H  =
0,\,j^E   = 0$  для  $x  \in  D$, то получим аналог второго закона
Ньютона  для твердого тела:
\[
\kappa \sqrt \varepsilon  \;\partial _\tau  \int\limits_{D^  -   }
{j^H  (x,t)dV(x)}  = q_{D^ -  } \tilde E' + m_{D^ -  } \tilde  H'
+ \int\limits_{D^  -   } {\left( {j^E  \times B'  -  j^H   \times
D'} \right)} \,dV(x).
\]

Если  (\ref{(4.7)})   скалярно  умножить на  нормаль  $n$  к
поверхности  $S$ и проинтегрировать по $S$ , то получим
                                 $$
 \kappa \sqrt \mu  \int\limits_S {\left( {\partial _\tau  j^E ,n(x)}
 \right)dS(x)}  + \kappa \varepsilon ^{ - 0,5} \int\limits_S {\left(
   {grad\rho ^E ,n(x)} \right)} \,dS(x) - $$
   $$- \kappa \sqrt \varepsilon \int\limits_l {\, \left( {\, j^H ,e(x)}
                            \right)dl(x)}=$$
                            $$
  = \int\limits_S {\left\{ {\left( {\rho ^E B' - \rho ^H D'} \right)
    + c^{ - 1} \left( {E' \times j^E  + H' \times j^H } \right)}
                          \right\}} \;dS(x).
                                 $$
Формула определяет изменение силы электрического тока через
заданную поверхность под воздействием  A-поля, порождаемого
другими зарядами и токами. Ее также можно проверить экспериментально.

\textit{Первое  начало  термодинамики.} Аналогично плотности
энергии  А- поля введем плотность энергии поля зарядов-токов:
\begin{equation}\label{(4.9)}
 0,5{\bf \Theta } \circ {\bf \Theta }^*  = \left( {
{\frac{\left\|{\rho _E }\right\|^2}{\varepsilon }}   +
{\frac{\left\|{\rho _H  }\right\|^2 }{\mu }}  + Q} \right) +
i\left( {P_J  - \sqrt {\frac{\mu }{\varepsilon }} \rho ^E j^E  -
\sqrt {\frac{\varepsilon}{\mu }} \rho ^H j^H } \right),
\end{equation}
которая содержит плотность энергии токов:
                                 \[
     Q  = 0,5\left\| J \right\|^2  = 0,5\left( {\mu \left\| {j^E }
   \right\|^2  + \varepsilon \left\| {j^H } \right\|^2 } \right),
                                 \]
где  первое  слагаемое   включает джоулево  тепло  $\left\|  {j^E
} \right\|^2  $,  а  второе - плотность кинетической энергии
массовых токов  $\left\| {j^H } \right\|^2 $, но не только, т.к. в
них входит и энергия вихревой части токов (магнитных токов). Здесь
также введен вектор $P_J  $, подобный вектору Пойнтинга,  но  для
токов:
                                 \[
      P_J  = 0,5i\,J \times \bar J =  c^{ - 2} \left[ {j^H ,j^E } \right]
                                 \]
Если гравимагнитный  и электрический токи параллельны, либо один
из них  отсутствует (нулевой), то $P_J  = 0$.  В общем случае $P_J
\ne 0$.

Умножим  скалярно  уравнение (\ref{(4.5)})  на  $  -  iJ^*  $,
сложим  с соответствующим комплексно-сопряженным и поделим на 2. В
результате получим

\textit{Закон сохранения энергии токов $\bf{\Theta}$-поля:}
\begin{equation}\label{(4.10)}
 \kappa \left( {\partial _\tau  Q  + \,div\,P_J  + {\mathop{\rm
     Re}\nolimits} \left( {\nabla \rho ,J^* } \right)} \right) =
  {\mathop{\rm Im}\nolimits} \left( {F,J^* } \right) = \left( {F^H
             ,J^H } \right) + \left( {F^E ,J^E } \right) ,
 \end{equation}
аналогичный  закону  сохранения энергии для А-поля  (теорема
1.1.). Однако  в левой части появилось третье слагаемое.  Нетрудно
видеть, что  этот  закон подобен первому началу термодинамики.
Здесь  сумма второго и третьего члена слева:
                                 \[
      U= \left( {\,div\,P_J  + \sqrt {\varepsilon /\mu } \left(
  {\nabla \rho ^E ,j^E } \right) + \sqrt {\mu /\varepsilon } \left(
               {\nabla \rho ^H ,j^H } \right)} \right),
                                \] --
характеризуют собственную скорость изменения плотности энергии
токов $\bf{\Theta}$-поля. Правая часть, зависящая от мощности
действующих внешних  сил, может увеличивать или уменьшать эту
скорость.

\vspace{3mm}
 \textbf{ 5. Свободное поле. Первый закон Ньютона.}
Рассмотрим А-поле в отсутствии других полей.  Назовем  такое поле
\textit{свободным}.

Аналогом  первого закона Ньютона об инерции массы  в  отсутствии
действующих  на  нее  сил здесь, как следует из  (\ref{(4.1)}) ,
естественно принять

\textit{    Закон инерции для зарядов-токов А-поля:}  В отсутствии
других полей $F=0$, т.е.
                                 \[
  {\bf D}^ -  {\bf \Theta } = \left( {\partial _\tau   - i\nabla }
                   \right){\bf \Theta } = {\bf 0}
                                 \]
что эквивалентно равенствам:
\begin{equation}\label{(5.1)}
{\partial _\tau  \rho  + div\,J}  = 0,\,\,\,\,
     \nabla \rho  + \partial _\tau  J - i\,\, rot\,J = 0                 .
\end{equation}
или для исходных величин:
\begin{equation}\label{(5.2)}
  \partial _t \rho ^E  + div\,j^E  = 0,\quad \partial _\tau  j^E
          = \sqrt {\varepsilon /\mu }\, rot\,j^H  - c\,grad\,\rho ^E,
\end{equation}
\begin{equation}\label{(5.3)}
\partial _t \rho ^H  + div\,j^H  = 0,\quad \partial _\tau  j^H
        =  - \sqrt {\mu /\varepsilon }\, rot\,j^E  - c\,grad\,\rho ^H
\end{equation}
Первые  равенства - это  законы сохранения  электрического заряда
и массы, из которых следует, что плотность массы изменяется только
под действием потенциальной составляющей токов. Из вторых
уравнений следует, что  градиенты зарядов  вызывают появление
соответствующих токов, кроме того вихревая гравимагнитная
составляющая тока влияет на электрический ток и наоборот.

Уравнения (\ref{(5.2)}) -(\ref{(5.3)})  являются замкнутой
системой уравнений для определения  плотностей  электрических
зарядов  и  токов, плотности массы и массовых токов А-поля в
отсутствии других полей.

Если  во втором уравнении (\ref{(5.1)})  взять дивергенцию или
ротор,  с учетом  уравнения для заряда и тока,  то получим
однородные волновые уравнения:  $\Box \rho   =  0,\quad \Box J =
0$. Что говорит  о том,  что  в свободном поле заряды и токи со
временем рассеиваются. В силу этого первое начало термодинамики
для свободного поля имеет вид:
                                 \[
  \partial _{\tau} Q  =-U, \quad U \geq 0.
                                 \]
или в интегральной форме:
                                 $$
\int\limits_{D^ -  } {\left( {Q (x,t) - Q (x,0)} \right)}
    dV(x) + \int\limits_0^t {dt} \int\limits_D {\,\left( {P_J ,n}
                       \right)} \,dD(x) +$$
                       $$
    + \int\limits_0^t {dt} \int\limits_{D^ -  } {\left( {\mu ^{ - 1}
      \left( {\nabla \rho ^E ,j^E } \right)} \right.}  + \left.
    {\varepsilon ^{ - 1} \left( {\nabla \rho ^H ,j^H } \right)\,}
                        \right)dV(x) = 0 .
                                $$

 \textit{Ударные волны. }Система уравнений (\ref{(5.1)})
гиперболического типа. Ее характеристические корни - нули
определителя
                                 \[
                    \left| {\begin{array}{*{20}c}
              { - \lambda } & {m_1 } & {m_2 } & {m_3 }  \\
          {m_1 } & { - \lambda } & { - im_3 } & { + im_2 }  \\
           {m_2 } & {im_3 } & { - \lambda } & { - im_1 }  \\
           {m_3 } & { - im_2 } & {im_1 } & { - \lambda }  \\
                      \end{array}} \right| = 0
                                 \]
определяют  скорость  распространения  ударных  волн.  В  данном
случае  $\lambda _{1,2}  =  \pm 1,\;$ т.е. скорость ударных волн
$\bf{\Theta}$-поля такая же, как и для А-поля. Два других корня
$\lambda _{3,4} = 0,$    т.е.,   как  и  в  случае уравнений
Максвелла, неподвижные поверхности в $R^3 $ являются
характеристическими [2].

Условия на фронтах ударных волн $F_\tau  $  легко получить, если
рассматривать  решения  уравнений  (\ref{(5.1)})    с разрывами на
характеристических поверхностях  как обобщенные решения (пометим
их ''шапочкой'': $\hat f$) :
                                                                  \[
 \partial _\tau  \hat \rho  + div\,\hat J = (\nu _\tau  \left[ \rho
     \right]_F  + \left( {\left[ J \right]_F ,\nu } \right))\delta _F(x,\tau ) =
     0,
                 \]
                                 \[
 \nabla \hat \rho  + \partial _\tau  \hat J - i\,\, rot\hat
 J = (\nu \left[ \rho  \right]_F  + \nu _\tau  \left[ J \right]_F  -
    i\,\left[ {\, \nu ,\left[ J \right]_F } \right])\delta _F(x,\tau) = 0.
             \]
Здесь  $(\nu _\tau  ,\nu _1 ,\nu _2 ,\nu _3 ) = ( - \left\|  \nu
\right\|,\nu ),$ нормальный вектор к характеристической
поверхности в  $R^4  $ ($\left\| \nu  \right\| = \sqrt {\nu _1^2 +
\nu _2^2   + \nu  _3^2  } $, связанный с волновым вектором в $R^3
$ соотношением: $m = \nu /\left\| \nu  \right\|$, $ \delta _F
(x,\tau )$-- сингулярная обобщенная функция - простой слой на $F$.
Следовательно, условия на фронтах ударных волн $J$-поля имеют
следующий вид:
\begin{equation}\label{(5.4)}
\left[ \rho  \right]_{F_\tau  }  - \left( {\left[ J
 \right]_{F_\tau  } ,m} \right) = 0,
\end{equation}
\begin{equation}\label{(5.5)}
 m\left[ \rho  \right]_{F_\tau  }  - \left[ J \right]_{F_\tau  }
- i\,\left[ {\, m,\left[ J \right]_{F_\tau  } } \right] = 0.
\end{equation}
Поскольку определитель этой системы уравнений равен нулю, она
имеет ненулевые решения. Первое условие такое же, как и для
ударных волн А-поля, связывает скачок объемной плотности
электрических зарядов и массовой плотности со скачком продольной
составляющей токов  на фронте волны. Второе условие дает связь
между касательными составляющими токов, которая, как следует из
(\ref{(5.4)}), не зависит от $[\rho]_{F_t}$.

Следовательно, в отличие от ударных волн напряжений А- поля,
которые являются поперечными, ударные волны $\bf{\Theta}$-поля,
таковыми, вообще говоря,  не являются. При отсутствии скачка
плотности заряда, эти волны становятся поперечными. \vspace{3mm}

\textbf{6. Уравнения единого поля и энергия взаимодействий}. Если
есть несколько (M) взаимодействующих полей, порождаемых различными
зарядами и токами, то уравнения (\ref{(5.1)}) примут вид:
\begin{equation}\label{(6.1)}
\kappa {\bf D}^ -  {\bf \Theta }^k  + {\bf \Theta }^k  \circ
                       \sum\limits_{m \ne k} {{\bf A}^m }  = {\bf 0}
                                                           ,\,\,\,\,
             {\bf D}^ +  {\bf A}^k  + {\bf \Theta }^k  = {\bf 0},k =
                                                       1,...,{\rm M}
\end{equation}
\begin{equation}\label{(6.2)}
{\bf D}^ +  {\bf A}^m  \circ {\bf A}^k  + {\bf D}^ +  {\bf A}^k
\circ {\bf A}^m  = 0,k \ne m.
\end{equation}
Суммарное (\textit{единое}) поле, в  силу  третьего закона Ньютона
(\ref{(4.4)})  для полей, как  легко видеть (суммируя
(\ref{(6.1)}) по  $k$), является свободным, поскольку, аналогично
механике  взаимодействующих тел, все действующие  силы внутренние.
Итак  имеем

\textit{ Закон   единого   поля.}  Взаимодействующие  поля
удовлетворяют аналогу второго закона  Ньютона для полей
(\ref{(4.4)}), (\ref{(4.5)}),  а для суммарного заряда-тока
выполняется тождество: $\kappa {\bf D}^ - {\bf \Theta } =  {\bf
D}^ - \sum\limits_{m = 1}^M {{\bf \Theta }^m }  \equiv {\bf 0}$ ,
т.е. единое поле является свободным.

Уравнения  суммарного  поля являются следствием  и  дают  первые
интегралы  взаимодействующих полей. Если в начальный момент
времени заряды-токи  известны,  система  (\ref{(6.1)})
-(\ref{(6.2)}) позволяет определять создаваемые  ими  поля  и  их
совместное изменение во времени  и пространстве.

Рассмотрим  законы  преобразования  энергии полей  при
взаимодействии различных зарядов-токов. Согласно (\ref{(3.1)}) ,
для  единого поля имеем
\begin{equation}\label{(6.3)}
  {\bf \Xi } = 0,5\,{\bf A}^*  \circ {\bf A} = 0,5\,(\bar A,A) -
      0,5\,[A,\bar A] = W + i\,P.
\end{equation}
Раскрывая, получим
$$
 {\bf \Xi }   = 0,5\sum\limits_{k = 1}^N
 {{\bf A}^k }  \circ \sum\limits_{l = 1}^N {{\bf A}^{*l} }  =
0,5\sum\limits_{k = 1}^N {{\bf A}^k  \circ {\bf A}^{*k} } +$$
\begin{equation}\label{(6.4)}
+ 0,5\sum\limits_{k \ne l} {{\bf A}^k  \circ {\bf A}^{*l} }
 =  \sum\limits_{k = 1}^N {{\bf \Xi }^k }  + {\bf \delta
}{\bf \Xi
  },
\end{equation}
\begin{equation}\label{(6.5)}
{\bf \delta }{\bf \Xi } = \sum\limits_{k \ne l}^{} {{\bf \Xi
}^{kl}  },\quad {\bf \Xi }^{kl}  = 0,5\left( {{\bf A}^k  \circ
{\bf A}^{*l}+  {\bf A}^l  \circ {\bf A}^{*k} } \right).
\end{equation}

Назовем  ${\bf  \delta  }{\bf  \Xi }$ бикватернионом
\textit{энергии-импульса }взаимодействия: $ {\bf \delta }{\bf \Xi
} = \delta W + i\delta P$. Здесь $\delta W$  - энергия
взаимодействия, $\delta P$- импульс взаимодействия, связанный с
изменением вектора Пойнтинга. Помечая скалярную и векторную часть
этого равенства соответствующими индексами, имеем $ {\bf \delta
}{\bf \Xi } =   \sum\limits_{k \ne l} {\left( {\delta W^{kl}  +
i\delta P^{kl} }\right)}$. Из    этих    соотношений   следует
представление   для \textit{энергии взаимодействия}:$  \delta W =
\sum\limits_{k \ne l}^{} {\delta W^{kl} }  .$

Поскольку  энергия  единого поля W - действительная
неотрицательная величина,  из   (\ref{(6.4)}) следуют условия на
выделение или   поглощение энергии при взаимодействии  А-полей:
выделение энергии, если $\delta W > 0,$ поглощение энергии, если
$\delta W  < 0$. При сохранении энергии-импульса  ${\bf \delta \Xi
}= 0$.

Закон сохранения энергии единого поля имеет вид, данный в теореме
1.1.

Заметим,  что  уравнения взаимодействия А-полей содержат  только
одну  универсальную размерную константу $\kappa$ , что позволяет
суммировать напряженности, заряды и токи и вводить единое поле.
При  переходе  к исходным    данным    можно    брать    константы
"проницаемости" $\varepsilon,\mu$   для   каждого   поля
взаимодействий   разными, математическая  теория  это  позволяет.
Частные  случаи   А-полей рассмотрены в [7].

\textbf{7. Заключение. }Предложенные здесь  уравнения
взаимодействия A-полей,  основаны на гипотезе о магнитном
заряде-массе, симметрирующем уравнения Максвелла. Это позволило
назвать такие поля \textit{электро-грави\-маг\-нит\-ными} и
построить законы их преобразования и взаимодействия, во многом
аналогичные законам Ньютона для материальных тел. Поскольку
квадрат размерности А-поля ($\alpha^2$)  - это размерность
плотности энергии, его можно назвать \textit{энергетическим}.

На основе этих законов можно объяснить ряд наблюдаемых физических
явлений. Например, хорошо известно, что  постоянные вращающиеся
электрические токи порождают соленоидальные магнитные поля [8]. Из
представленной теории следует (см. примеры в [1]), что аналогично,
вращающиеся массы создают соленоидальные электрические поля (что
можно экспериментально проверить). Это объясняет, в частности,
наличие электрической оси у Земли, обусловленной соленоидальной
составляющей электрического поля Земли вследствие вращения ее
массы. Подобное должно наблюдаться и у других планет. Поскольку
Земля имеет отрицательный электрический заряд, то его вращение
обуславливает и  соленоидальное магнитное поле Земли.

Другой пример...  Согласно этой теории следует учитывать, что
электрическое поле Солнца порождает  электромасовую силу,
смещающую орбиты планет. Интересно, что в работе [9] для
определения смещения перигелия Меркурия, необъяснимого в рамках
классической теории гравитации,  в уравнения Ньютона для двух тел
введена дополнительная сила, названная авторами
\textit{когравитацией}, которая по виду подобна рассмотренной
здесь электромассовой силе.

Представленная теория дает  также уравнения  изменения
электрических токов и зарядов под действием электро-гравимагнитных
полей и законы преобразования энергии-импульса, которые могут
описать, например, некоторые закономерности химических реакций
веществ.

\vspace{3mm}

Ключевые слова: \textit{ электро-гравимагнитное
поле,бикватернионы, напряженность, заряд-ток, мощность-сила,
уравнения взаимодействия,  свободное поле, ударные волны,
условия на фронтах, законы сохранения, энергия взаимодействия}
\vspace{3mm}

Keywords: \textit{ electro-gravymagnetical field,biquaternion, tension,
charge-current, power-force, interaction equations , free field,
shock waves, condition on front, conservation laws, energy of the interaction}
\vspace{5mm}

\centerline{  \textbf{Литература}}

1.  Алексеева Л.А. О замыкании уравнений Максвелла //  Журнал
    вычислительной математики и математической физики. -Т.43.- №5.-
    2003.- C.759-766.

2.   Алексеева  Л.А. Гамильтонова форма уравнений Максвелла  и  ее
     обобщенные решения// Дифференциальные уравнения. -Т.39.-№6.-2003.-
     С.769-776

3.  Алексеева Л.А. Кватернионы гамильтоновой формы уравнений
    Максвелла// Математический журнал. -Алма-Ата.-2003.-Т.3.-№4.-С.20-24

4.   Шнеерсон М.С.О моногенных функциях Мойса//Математический сборник Т.44(86)
     №1.С113-122

5.   Казанова Г. Векторная алгебра.М.:Мир.1979.120с.

6.   Kassandrov V.V.Buquaternion electroynamics and Weyl-Cartan geometry of space-time//
     Gravitation and cosmology.V.1(1995).№3.Pp.216-222

7.   Алексеева Л.А. Об одной модели электро-гравимагнитного  поля.
     Уравнения  взаимодействия и законы сохранения//  Математический
     журнал.-Алма-Ата.-2004.-Т.4.-№2.- С.23-34.

8.   Тамм И.Е. Основы теории электричества. -М.-1989.

9.   Matos C.J., Tajmar M. Advance of Mercury Perihelion Explained
     by Co\-gra\-vi\-ty/ Advanced Concepts and Studies Officer.2003. E-mail:
     clovis.de.matos@esa.int

\end{document}